\documentclass[aps,prl,twocolumn,superscriptaddress]{revtex4}
\usepackage{amsmath,epsfig,mathrsfs,graphicx,amssymb}

\begin{document}

\title{Formulation of Time-Resolved Counting Statistics Based on a Positive-Operator-Valued Measure}

\author{Adam Bednorz}
\email[]{Adam.Bednorz@fuw.edu.pl}
\affiliation{Fachbereich Physik, Universit{\"a}t Konstanz, D-78457 Konstanz, Germany}
\affiliation{University of Warsaw, Ho\.za 69, PL-00681 Warsaw, Poland}
\author{Wolfgang Belzig}
\affiliation{Fachbereich Physik, Universit{\"a}t Konstanz, D-78457 Konstanz, Germany}
\pacs{73.23.-b, 72.70.+m}

\date{\today}
\begin{abstract}
  We propose a derivation of the full counting statistics of
  electronic current based on a positive-operator-valued measure. Our
  approach justifies the Levitov-Lesovik formula in the long-time
  limit, but can be generalized to the detection of finite-frequency
  noise correlations. The combined action of the projection postulate
  and the quantum formula for current noise at high frequencies imply
  an additional white noise. Estimates for this additional noise are in
  accordance with known experiments. We propose an experimental test
  of our conjecture by a simultaneous measurement of high- and
  low-frequency noise.
\end{abstract}
\maketitle

The core of quantum measurement theory is the projection postulate
\cite{neumann}.  It provides a consistent description of a sequence of
measurements. Quantities represented by non-commuting operators cannot
be measured simultaneously. The corresponding projection operators
have to be time-ordered. For continuous variables the projection postulate should be
replaced by a positive-operator-valued measure (POVM) \cite{hpp}. The
idea of the POVM is that one does not measure the exact value for a
given operator but a finite accuracy is taken into account due to some
interaction with the detector and its internal dynamics. However, due
to Naimark's theorem \cite{naimark}, every POVM can be realized by a
set of orthogonal projections in an extended Hilbert space. The
resulting POVM will depend on the detection scheme.

The statistical behavior of current flow in a quantum point
contact 
can be found by measurement of the correlation functions. The long-time cumulants of the transferred charge can be derived from the
Levitov-Lesovik formula \cite{lev1}, which led to the foundation of
the electronic version of full counting statistics (FCS)
\cite{bel,naz1,naz2}. It has been confirmed experimentally for noise
\cite{glatt1} and third cumulant \cite{reulet,reznikov, others}.  On
the other hand, the current-current correlation function (noise
spectral density) is given by the quantum noise
\cite{blanter,lesovik}, which coincides with the FCS result at low
frequencies.  The high-frequency quantum noise can be obtained by a
generalization of FCS to finite frequencies with additional
predictions for higher cumulants at arbitrary frequency
\cite{zaikin1,salo}. Also the semiclassical predictions of the third
cumulant are consistent with purely quantum results in some limits
\cite{pil}.  The behavior of quantum noise has been confirmed
experimentally also for high frequencies
\cite{schoel,glatt2,gabel}. From the fundamental point of view, the
low-frequency results can be justified by a proper use of the
projection postulate but there is no unique derivation for high
frequencies \cite{lesovik08}. A similar problem occurs for a chain of
spin-resolved detectors, for which the results depend on the detector
properties \cite{lorenzo}. While it is reasonable to expect an
influence of the detector on the outcome \cite{naz2,lorenzo,martin},
it should be possible to separate it from the bare signals of the
sample.

In this Letter we address the question, if a general definition of FCS
for finite frequencies is possible - maintaining the probabilistic
interpretation. We will demonstrate that the standard definition of
FCS, when generalized to finite frequency, can lead to negative
probabilities. To cure this deficiency, we show that taking into
account a minimal model of a detector, a POVM of FCS can be introduced,
which leads to positive definite probabilities. 



The definition of the generating functional for a probability distribution $\varrho$ of
a given time trace of the current through a quantum point contact, $I(t)$, is
\begin{equation}
e^{S[\chi]}=\langle e^{\int i\chi(t)I(t)dt/e}\rangle_\varrho=
\int DI\;\varrho[I]e^{\int i\chi(t)I(t)dt/e}.\label{defg}
\end{equation}
On the other hand, one can first define FCS generating function  
\cite{lev1,naz1,naz2,zaikin1}
\begin{equation}
e^{S[\chi,\phi]}=\mathrm{Tr}\hat{\rho}\tilde{\mathcal T}e^{\int
  \frac{idt}{2e}[\chi(t)+2\phi(t)]\hat{I}(t)} 
\mathcal Te^{\int \frac{idt}{2e}[\chi(t)-2\phi(t)]\hat{I}(t)}.\label{genf1}
\end{equation}
Here $\hat\rho$ denotes initial state density matrix, $\hat{I}(t)$ is
the Heisenberg current operator, $\phi$ refers to classical phase
bias, and $\mathcal{T}(\tilde{\mathcal{T}})$ denotes (anti-)time
ordering. A detailed definition of $\hat{I}(t)$ will be given later.
 
Taking $S[\chi]=S[\chi,0]$ we obtain $\varrho$ by inverse Fourier transform of (\ref{defg}).
However, this gives positive probabilities only in the zero-frequency limit. For time-dependent
quantities, we can construct the following counterexample for a
single mode point contact at zero temperature in the tunneling limit
(transmission $T\ll 1$). Let us define
\begin{equation}
  X=\int_0^{t_0}dtdt'I(t)I(t') 
  [e^{-(t-t')^2/s^2}-2e^{-(t-t')^2/9s^2}].
  \label{xxx}
\end{equation}
Then, following \cite{zaikin1} we find $\langle (\delta X)^2
\rangle_\varrho=-Tt_0e^4/3s\pi^{3/2}$ for $\delta X=X-\langle X\rangle_\varrho$ and
 $t_0\gg s$. This obviously contradicts the interpretation of
$\varrho$ as probability.

To overcome this fundamental problem, we now construct a positive
definite probability of time-dependent FCS based on a POVM.  Instead
of the projection operator we define the more general Kraus
operator \cite{kraus}
\begin{equation}
  \hat{K}[I] = \int D\varphi \mathcal{T}e^{\int dt [i\varphi(t)[\hat{I}(t)-I(t)]/e
    -\varphi^2(t)/\tau]}\;.\label{krau}
\end{equation}
Causality is preserved since the detector affects the measurement only
in later times.  The time scale $\tau$ describes internal fluctuations
of the detector and depends on its temperature in general. For
$\tau\to\infty$ the measurement is accurate, but the detector noise
strongly affects the system by full projection. A shorter $\tau$
reduces the influence of detector but induces a larger measurement
error. The integration measure contains also a normalization factor to
be determined later.  The positive definite probability of a given
$I(t)$ is defined as
\begin{equation} 
  \rho[I]=\mathrm{Tr}\;\hat{\rho}
\hat{K}^\dag[I]\hat{K}[I], \label{rdef}
\end{equation}
for the given initial density matrix $\hat{\rho}$. We note that our
choice of the Kraus operator represents generically the
influence of a detector, parametrized by a single parameter
$\tau$. What concrete models of detectors lead to our definition of
the Kraus operator is an interesting question, which we will not
address here.

We now substitute in Eq.~(\ref{rdef}) $\varphi\to\phi\pm\chi/2$ in
$\hat{K}$ and $\hat{K}^\dag$, respectively.  The generating functional
${\mathcal S[\chi]} = \ln\langle \exp(i\int
dt\chi(t)I(t)/e)\rangle_\rho$ necessary for the calculation cumulants
takes the form
\begin{equation}
  \mathcal S[\chi]
  =\ln\int D\phi\: e^{S[\chi,\phi]-\int dt[2\phi^2(t)+\chi^2(t)/2]/\tau}\;,\label{genf}
\end{equation}
where $S[\chi,\phi]$ is defined by (\ref{genf1}).  The measure $D\phi$ is
scaled to keep $\mathcal S[\chi\equiv 0]=0$.  The measuring device
affects the generating function by the additional exponent in
(\ref{genf}). In \cite{lorenzo} Di Lorenzo and Nazarov used the expression
$\tau^2\dot\phi^2+\phi^2/(\Delta \phi)^2$ instead of $\phi^2$, with
$\Delta\phi$ as an additional parameter, and considered low-frequency measurements. 
In contrast, we rather assume a continuous weak measurement of the system to obtain 
finite frequency correlations.

\begin{figure}[t]
\includegraphics[scale=.4]{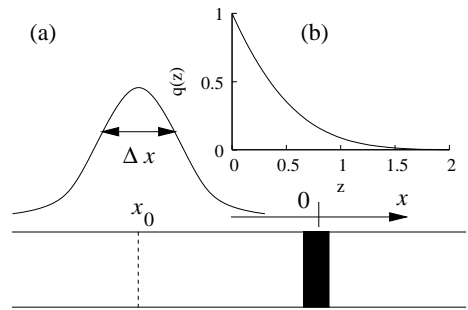}
\caption{(a) Sensitivity of current measurement. The Gaussian
  distribution refers to the dispersion of current measurement. (b)
  The function $q(z)$.}
\label{ff1}
\end{figure}

To further model our measuring device, we note that in general a
current measurement has also a \emph{spatial} sensitivity, i.e. the
point of the measurement is not exact. In experiments, it can be
usually related to the finite capacitance of the sample.  Therefore,
we assume a generic form of the current operator in a
quasi-one-dimensional lead as
\begin{equation}
\hat{I}(t)=\int \hat{I}(x,t)  e^{-\frac{(x-x_0)^2}{2\Delta x^2}}dx/(\sqrt{2\pi}\Delta x)\,.
\end{equation}
The setup is shown in Fig.~\ref{ff1}a. The real dispersion may be
non-Gaussian. However, we stress that our model is general enough to
capture the essential physics, but still allows some analytical
progress.

We will assume non-interacting electrons and energy- and
spin-independent transmission through the $M$ mode junction. We count
\emph{all} modes, although most of them are just reflected and denote
the Fermi velocity, the transmission and the reflection coefficients
for mode $n$ by $v_n$, $T_n$ and $R_n=1-T_n$, respectively.  For
convenience, we introduce $t_n=|x_0|/v_n$ and $\tau_n=\Delta
x/v_n$. The times $\tau_n$ are related to $RC$ times of the circuit,
which limits the observable frequencies to $\omega \lesssim
\tau_n^{-1}$.  Furthermore, we assume that $t_n^{-1} \ll
\tau_n^{-1}$, which means that the detector sensitivity function is
entirely located on one side of the junction. 

To model the electron transport we apply the standard scattering
picture around the Fermi level \cite{blanter}.  The Hamiltonian can be
approximated by $\hat{H}=\sum_{\bar n}\int dx\hat{\mathcal H}_{\bar
  n}(x)$, where
\begin{eqnarray}
  \hat{\mathcal H}_{\bar n}&=& 
  i\hbar v_n [\hat{\psi}^\dag_{L\bar n}(x)\partial_x\hat{\psi}_{L\bar
    n}(x)-\hat{\psi}^\dag_{R\bar n}(x)\partial_x\hat{\psi}_{R\bar
    n}(x)]\\\nonumber 
  &&+q_n\delta(x)[\hat{\psi}^\dag_{L\bar
    n}(x)\hat{\psi}_{R\bar n}(x)+\hat{\psi}^\dag_{R\bar
    n}(x)\hat{\psi}_{L\bar n}(x)]\\
&&-eV\theta(x)[\hat{\psi}^\dag_{L\bar
    n}(x)\hat{\psi}_{L\bar n}(x)+\hat{\psi}^\dag_{R\bar
    n}(x)\hat{\psi}_{R\bar n}(x)].\nonumber 
\end{eqnarray}
The scattering states obey standard fermionic anticommutation relations
$\{\hat{\psi}^\dag_{A\bar n}(x),\hat{\psi}_{B\bar
  m}(y)\}=\delta_{AB}\delta_{\bar n\bar m}\delta(x-y)$ and
$\{\hat{\psi}_{A\bar n}(x),\hat{\psi}_{B\bar m}(y)\}=0$.  Here $A=L,R$
denote left and right going state, $\bar n=(n,\sigma)$ denotes mode
number $n$ and spin orientation $\sigma$. The transmission coefficient
is given by $T_n=\mathrm{cosh}^{-2}(q_n/\hbar v_n)$.  The current
operator is defined as $\hat{I}(x) = \sum_{\bar n}e
v_n\hat{\psi}^\dag_{L\bar n}(x) \hat{\psi}_{L\bar n}(x)
-L\leftrightarrow R$. The initial density matrix for a thermal state
is $\hat{\rho}=e^{-\hat{H}/k_BT}/\mathrm{Tr}e^{-\hat{H}/k_BT}$ and
the time evolution is governed by the Heisenberg operator
$\hat{I}(x,t)=e^{i\hat{H}t/\hbar}\hat{I}(x) e^{-i\hat{H}t/\hbar}$.

For our model, the mean current is independent of the detector,
$\langle I(t)\rangle_\rho=GV$, where the conductance $G=\sum_n T_n G_Q$ and
$G_Q=e^2/\pi\hbar$.  We define the noise
spectral density as a second cumulant $ e^2P(\omega)=\int
dt\:e^{i\omega t}\langle\delta I(0)\delta I(t)\rangle_\rho$, where
$\delta I(t)=I(t)-\langle I(t)\rangle_\rho$.  
It is calculated from the functional derivative
\begin{equation}
\langle\delta I(0)\delta I(t)\rangle_\rho=-e^2\left.\frac{\delta^2 \mathcal S[\chi]}{\delta\chi(t)\delta\chi(0)}\right|_{\chi\equiv 0},
\end{equation}
where $\mathcal S[\chi]$ is defined by Eq.~(\ref{genf}).  In our
construction, the noise is a classical quantity and, hence, symmetric
with respect to $\omega$.  We consider frequencies $|\omega\tau_n|\ll
1$, since we do not include capacitive effects and obtain
\begin{equation}
P(\omega)=P_{\textrm{off}}+P_{S}(\omega)+P_0(\omega)+P_\tau(\omega)
+P_\Delta(\omega).
\end{equation}
Let us discuss the behavior of all terms of this expression.  The
first one, $P_{\textrm{off}}=1/\tau$ is a white offset noise,
independent of temperature and voltage bias.  Defining
$w(\omega)=\omega\mathrm{cth} (\hbar\omega/2k_BT)$ and
$w_\pm(\omega)=w(\omega\pm eV/\hbar)$ the second term
\begin{equation}
P_S(\omega)=\sum_n\frac{T_n}{2\pi}\{2T_nw(\omega)+R_n[w_+(\omega)
+w_-(\omega)]\}\label{elb}
\end{equation}
is just the symmetrized quantum noise $\int dt \cos(\omega t)
\mathrm{Tr}\hat{\rho} \delta\hat{I}(0) \delta\hat{I}(t)/e^2$
\cite{blanter}. However, for energy independent transmission, as we
assume here, the asymmetric noise contains only the additional term
$\sum_nT_n\omega/\pi$, which is independent of temperature and
voltage. The next term is
$
P_0(\omega)=\sum_n2R_n\sin^2(\omega t_n)w(\omega)/\pi.
$
This is a contribution to the quantum noise due to the finite flight time
to the detector, as it depends on $t_n$. Note that it is independent of voltage and
sensitivity.  The problem of flight time has been already discussed in
context of third cumulant \cite{reznikov, chelk}, but there is no
experimental evidence of its influence on the noise. The detection noise,
\begin{equation}
  P_\tau(\omega)=\frac{\tau}{4\pi^2}\left|\sum_n\left[\omega(1+R_ne^{2i\omega 
    t_n})+\frac{i}{\sqrt{\pi}\tau_n}\right]\right|^2\label{taaa}
\end{equation}
combines the effects of the measurement sensitivity $\tau$ and flight times
$t_n$ but is independent of voltage and temperature. Finally,
\begin{equation}
  P_\Delta(\omega)=\int
  \frac{d\alpha}{(2\pi)^{2}}\sum_n f_n(\omega-\alpha)R_nT_n[w_+(\alpha)+
  w_-(\alpha)]\label{ede} 
\end{equation}
is an additional mixed noise. Here the sensitivity amplitude is given by
\begin{equation}
f_n(\alpha)=\int
dt\;\left[\exp\left(-\frac{1-e^{-t^2/4\tau^2_n}}{8\sqrt{\pi}\tau_n/\tau}
  \right)-1\right]e^{i\alpha t}.  
\end{equation}
It is independent of the flight times, but all other parameters enter
in a rather complicated way.  Eq.~(\ref{ede}) is the only term
depending on $\tau_n$ in the limit $|\omega\tau_n|\ll 1$ However,
voltage and temperature are arbitrary.  For $\tau\ll\tau_n$ the
amplitude reduces to $f_n(\alpha) = \tau(e^{-\alpha^2\tau_n^2} -
\delta(\alpha)\sqrt{\pi}/\tau_n)/4$.

We assume that most modes are closed, $\sum_n T_n\ll M$, which is true
in many experimental setups (e.~g., tunnel barrier, diffusive wire,
quantum point contact). We will consider several interesting limits:
short and long wire (flight time), low and high frequencies and zero
temperature. They correspond to the most common experimental setups.
For short flight times, $|\omega t_n|\ll 1$, we have $P_0 =
2\omega^2w(\omega) \sum_n t^2_n$. For long flight times, $|\omega t_n|\gg 1$, we
get $P_0=Mw(\omega)/\pi$
because random flight times imply $\sin^2(\omega t_n)\to 1/2$.
In both cases $P_\tau=(\sum_n\tau_n^{-1})^2\tau/4\pi^3$
is independent of voltage and yields additional offset noise.  For low frequency
and a slow detector, $\hbar|\omega|\ll (\hbar/\tau , \hbar/\tau_n)\ll
eV,k_BT$ the mixed noise is negligible since $P_\Delta\ll
P_{S}(0)$. However, for $k_BT\ll\hbar|\omega| \ll
\hbar/\tau_n\ll\hbar/\tau$, we have $ P_\Delta=(\tau/8\pi^2)
\sum_nR_nT_nq(\tau_n|eV|/\hbar)/\tau^2_n $, where
$q(z)=e^{-z^2}-2z\int_z^\infty e^{-t^2}dt$. The decay of $q(z)$ is shown in
Fig.\ref{ff1}b.  It vanishes for $eV\gg\hbar/\tau_n$, which means that
the size of the wave packet becomes smaller than the spatial
sensitivity of the detector.

From the above results we conclude, that the POVM reproduces the
standard quantum result in the case $|\omega| \ll
(1/\tau_n,1/t_n)\ll 1/\tau \ll eV/\hbar , k_BT/\hbar$. It can be
shown, that corrections to higher zero-frequency cumulants are
negligible in this case, because $P_\Delta$ is small. Hence, the
Levitov-Lesovik formula and FCS \cite{lev1,naz1,bel,naz2} are
justified, as expected.

The situation becomes more interesting, if we look at the high
frequency quantum noise.  We can make $P_\tau$ and $P_\Delta$
negligible by choosing a very small $\tau$, which corresponds to a
weak detection. For small $t_n$, also $P_0$ gives only a small
contribution. Moreover, higher cumulants then have also negligible
corrections to predictions of generalized FCS as small $\tau$
corresponds to $\phi\to 0$ in eq. (\ref{genf1}). What remains is the
large white Gaussian offset noise $P_{\textrm{off}}$ -- the price we
have to pay for small $\tau$.  Conversely, lowering $P_{\textrm{off}}$
will increase $P_\Delta$, which additionally depends on voltage.  They
become of the same order at $\tau\sim\tau_n$ and $P_\Delta$ is growing
as $\sqrt{\tau/\tau_n}$ for $\tau\gg\tau_n$. Hence, one cannot get rid
of the additional noise by increasing $\tau$, since it increases the
backaction noise.

The offset noise for a few mode quantum point contact in most high
frequency experiments \cite{schoel,glatt2} was usually
subtracted. However, the results of the recent experiments
\cite{glatt2,gabel} show relatively high absolute noise temperature
$T^\ast=G_QPh/4Gk_B$, setting an upper bound to the offset-noise. We
find $T^\ast=T_{\textrm{off}}+T_\Delta$ with $T_{\textrm{off}}=24$K
and $T_\Delta=1.5$mK for a single-mode quantum point contact with
transmission $1/2$, $\tau=1$ps and $\tau_1=10$ps.  Although
$T_{\textrm{off}}$ is larger, it is constant whereas $T_\Delta$
depends on voltage bias and drops to zero according to the function
$q(z)$. For zero temperature this yields a characteristic voltage of
$130\mu$V at $z=1$.  Diffusive and tunnel barriers have usually a
higher conductance due to the large number of modes and hence the
offset temperature can be much lower. On the other hand, in a
Josephson junction or a quantum dot as a detector, the measured
quantity is more qualitatively than quantitatively related to the
frequency dependent quantum noise \cite{kouw}.

Moreover, as the offset noise is white, it must be visible also at low
frequencies.  It would be therefore of great interest to test high and
low frequency noise \emph{simultaneously}.  The high frequency
detector measures only the difference $P(eV)-P(0)$ whereas at low
frequencies the absolute noise is measured. The fluctuation-dissipation
theorem \cite{imgav} will of course be maintained after subtraction of the
offset noise and then taking the limit $\tau\to 0$.

We generalize our approach to a multi-terminal measurement. We define
the Kraus operator (\ref{krau}) as
\begin{equation}
\hat{K}[I]=\int D\phi\: \mathcal Te^{\int dt\sum_A\{i\phi_A(t)[\hat{I}_A(t)-I_A(t)]/e
-\phi^2_A(t)/\tau_A\}}\label{kra1}
\end{equation}
where $A$ labels the terminals. A simple example is the noise between
the left and right sides of a junction, namely $e^2P_{AB}(\omega) =
\int dt\:e^{i\omega t}\langle\delta I_A(0) \delta I_B(t)\rangle$ where
$A,B=L,R$ for left and right terminals (or equivalently $x_0<0$ and
$x_0>0$), respectively.  The cross correlation $P_{LR}(\omega)$ is
finite only for $|\omega t_n|\ll 1 $, since otherwise one averages
over different $t_n$.  In this limit, we have
\begin{eqnarray}
&&
P_{LL}=P_{L\textrm{off}}+P_{S}+P_{\tau_L}+P_{\Delta L}+P_{\Delta R},\nonumber\\
&&
P_{RR}=P_{R\textrm{off}}+P_{S}+P_{\tau_R}+P_{\Delta L}+P_{\Delta R},\nonumber\\ 
&& P_{LR}=P_{RL}=-P_{S}+P_{\textrm x}-P_{\Delta L}-P_{\Delta R},
\end{eqnarray}
where
$P_{\textrm x}(\omega)=\omega^2(\tau_L+\tau_R)M\sum_n T_n/4\pi^2$.
Here $P_{\textrm{off}A}=\tau_{A}^{-1}$, $P_{\tau_{A}}$ and $P_{\Delta
  A}$ are given by Eqs. (\ref{taaa}) and (\ref{ede}) with $\tau$ and
$\tau_n$ replaced by $\tau_{A}$ and $\tau_{nA}$.  $P_\Delta$ is here
replaced by the sum of contributions of both detectors.  The cross
noise does not contain the offset noise, so it can help to estimate
the measurement timescales, in particular $P_{\Delta L}+P_{\Delta R}$ at
low frequency. 

Finally, we propose the following test of our definition of a quantum
probability.  Consider two detectors, similar to \cite{reznikov},
measure $X_L$ and $X_R$ defined by (\ref{xxx}) for $I=I_L$ and $I_R$,
respectively.  The classical expectation $X_L\stackrel{cl}{=}X_R$
based on charge conservation in the low frequency limit, leads to
$\langle\delta X_L\delta X_R\rangle_{cl}>0$. The quantum measurement
using the probability density $\rho[I]$, results for $\tau\ll s$ in
$\langle\delta X_L\delta X_R\rangle_\rho=\langle(\delta
X)^2\rangle_\varrho$. This is, however, negative as we have shown in
beginning.

In conclusion, we have constructed a positive probability measure,
based on POVM, that justifies the use of  Levitov-Lesovik formula and
FCS in long time (low frequency) limit. Our approach cures certain
deficits in the standard definition of FCS, which lead to negative
probabilities. Introducing a generic one-parameter model of the
influence of the detector, we predict an intrinsic additional white
offset noise. Such a noise is in agreement with recent experiments and
a further verification by simultaneous low- and high-frequency noise
measurements would be desirable.

We acknowledge interesting discussions with C. Bruder and G. Burkard
and financial support from the German Research Foundation
(DFG) through SFB 513, SFB 767, and SP 1285
\textit{Semiconductor Spintronics}.

\end{document}